\documentclass{article}
\usepackage{spconf,amsmath,graphicx}
\usepackage{psfrag}
\usepackage[hidelinks]{hyperref}

\usepackage[dvipsnames]{xcolor}  
\newcommand\figref{Fig.~\ref}    
\usepackage{pgfplots}       
\pgfplotsset{compat=newest} 
\pgfplotsset{plot coordinates/math parser=false} 
\usetikzlibrary{plotmarks}  
\usepackage{import}         
\usepackage{cite}           

\title{On Neural-Network Representation of Wireless Self-Interference\\ for Inband Full-Duplex Communications}
\name{Gerald Enzner$^1$\thanks{This work is funded by the German Research Foundation (DFG) under Grant EN 869/4-1, Project ID 449601577.}, Aleksej Chinaev$^1$, Svantje Voit$^1$, Aydin Sezgin$^2$\vspace{-1ex}}
\address{$^1\,$Department of Medical Physics and Acoustics, Carl von Ossietzky Universität Oldenburg, Germany\\
$^2\,$Chair of Digital Communication Systems, Ruhr-Universität Bochum, Germany\\
Email: \{gerald.enzner, aleksej.chinaev\}@uni-oldenburg.de}
\begin{document}
\ninept
\maketitle
\begin{abstract}
Neural network modeling is a key technology of science and research and a platform for deployment of algorithms to systems. In wireless communications, system modeling plays a pivotal role for interference cancellation with specifically high requirements of accuracy regarding the elimination of self-interference in full-duplex relays. This paper hence investigates the potential of identification and representation of the self-interference channel by neural network architectures. The approach is promising for its ability to cope with nonlinear representations, but the variability of channel characteristics is a first obstacle in straightforward application of data-driven neural networks. We therefore propose architectures with a touch of "adaptivity" to accomplish a successful training. For reproducibility of results and further investigations with possibly stronger models and enhanced performance, we document and share our data.
\end{abstract}
\begin{keywords}
\vspace{-0.5ex}
system modeling, neural networks, full-duplex
\vspace{-0.5ex}
\end{keywords}

\section{Introduction}
\vspace{-1ex}
\label{sec:intro}

Inband full-duplex (IBFD) communication uses one and the same frequency resource for simultaneous sending and receiving. Hence, it requires cancellation of the large self-interference (SI) of the transmitter into the receiver unit. The technology status implies the possibilities of active or passive SI shielding in the propagation domain, the adaptive or non-adaptive cancellation in the analog receiver unit, and the adaptive cancellation in the digital baseband section of the receiver \cite{Heino_2015,Herd_2019,Smida2023}. The different treatments would need to be joined in an implementation of an IBFD system in order to cope with the huge requirement of about 100\,dB SI cancellation (SIC).

On the digital side, the majority of approaches has been concerned with signal processing for SI channel estimation and SIC according to maximum-likelihood, subspace \cite{Le-Ngoc_2017}, mean-square error \cite{Tepedelenlioglu_2018}, least-squares \cite{Guo_2023}, and linear \cite{Zhou_2023} or nonlinear adaptive-filter methods \cite{Vogt_2019}. Considering limitations w.r.t.\ the extraordinary accuracy required for SIC, another approach consists in the design of robust transceivers in the presence of SI at the receiver by non-convex optimization \cite{Sezgin2020} or modified matched filtering \cite{Mohammad_2023}. The transmitter side as well can be optimized to support the SIC task, e.g., by active injection of compensation signal \cite{Le-NgocTho2017FWCS} or by the choice of suitable pilot sequences with favorable nonlinear behavior \cite{Kong_2019}.

The omnipresent machine learning, specifically, a neural network representation for SIC has been pursued more rarely \cite{PANSE2022101526}. Neural networks demonstrated to line up or even improve over linear and polynomial modeling accuracy, while reducing computational complexity \cite{Zhang_2018,Guo2019,Stimming2019}. Typically, SI modeling separately represents the linear (radio channel) and nonlinear (power amplifier) elements \cite{Stimming2019,Baek2019,Muranov2021} in a very model-oriented and interpretable approach.

The somewhat rare use of neural network models in this field may be attributed to the limitation of trainable models in case of typically time-varying SI characteristics and in the limited availability of datasets for comparative analysis and conclusive benchmarking of the models \cite{PANSE2022101526}. Both of these limitations are addressed with the approach of our paper. We specifically demonstrate the failure of common network architectures for the desired SI modeling in case of variable linear or nonlinear elements in the data. As a remedy we therefore propose architectures with a blend of training and adaptation, i.e., with a subset of trainable weights to represent invariant and another subset of ''adaptive'' weights to fit variable characteristics of the data. Moreover, the data prepared for our analysis is documented and made publicly available for reproducibility and further utilization in the community. The data is synthetic and meant as baseline for development and analysis, notwithstanding the utilization of real SI recordings which is beyond the scope here.

Sec.~\ref{sec:system} of the paper depicts the full-duplex SI problem and two basic SI cancellation options (i.e., Wiener and Hammerstein based). Sec.~\ref{sec:data} describes the related construction of research data with varying levels of difficulty. Sec.~\ref{sec:architectures} then introduces neural network architectures for Hammerstein-based SI modeling with varying effort. Sec.~\ref{sec:results} relates the data and the models by results. Sec.~\ref{sec:conclusion} concludes. \vspace{-1ex}

\section{Overview of Full-Duplex System}
\vspace{-1ex}
\label{sec:system}

From the wide range of possible system options for SIC, we here consider two opposite system designs, a Wiener and a Hammerstein type of system, for each of which we provide the data. Development and analysis of neural network architectures for the data at hand, however, was currently limited to the Hammerstein subset.

\begin{figure}[!tb]
  \centering 
  \psfrag{f(s)}[][]{$\!f(s)$}
  \psfrag{g(x)}[][]{$g(x)$}
  \psfrag{w(k)}[][]{$w_k$}
  \psfrag{s(k)}[][]{$s[k]$}
  \psfrag{r(k)}[][]{$r[k]$}
  \psfrag{y(k)}[][]{$y_\text{W}[k]$}
  \psfrag{z(t)}[][]{$z(t)$}
  \psfrag{hSI}[l][l]{\,$h_\text{SI}(t)$}
  \psfrag{yH}[][]{$\!\!\!y_\text{H}(t)$}
  \psfrag{x(t)}[][]{$x(t)$}
  \psfrag{adaptive system}[l][l]{\small \,\,\,\,neural net}
  \includegraphics[width=\columnwidth]{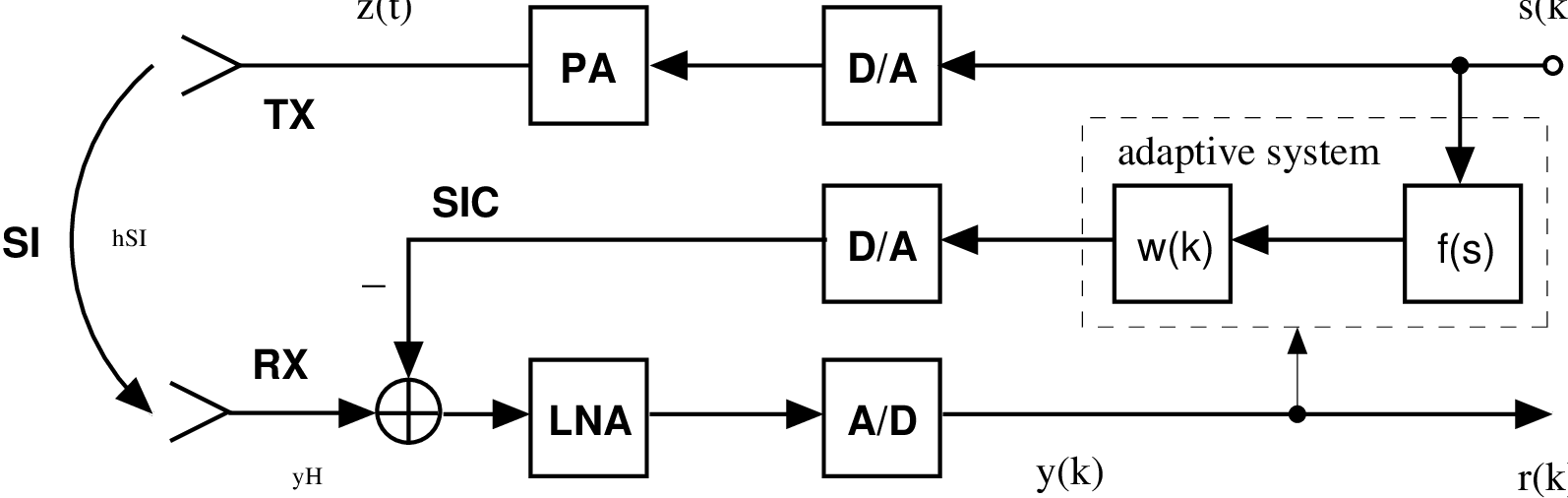}\\[1ex]
  a) system option with ''Hammerstein'' (H) network across D/A\\[3ex]
  \includegraphics[width=\columnwidth]{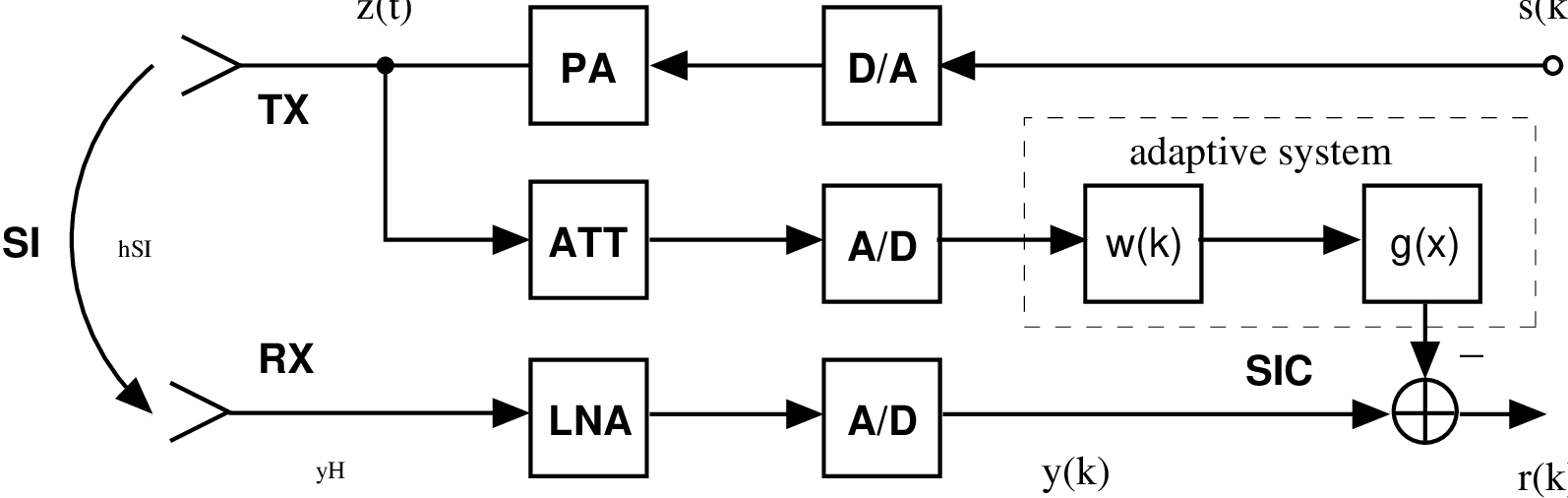}\\[1ex]
  b) system option with ''Wiener'' (W) type network across A/D\\[-0.5ex]
  \caption{System options with self-interference cancellation.}
  \vspace{-2.5ex}
  \label{fig:baseline}
\end{figure}

Fig.~\ref{fig:baseline}.a shows the baseband design with ''Hammerstein'' logic (i.e., generally a memoryless nonlinearity followed by a linear dynamical system element) for SIC. The transmit signal $s[k]$ at discrete time $k$ is D/A converted and passed to the TX antenna via a nonlinear power amplifier (PA). The wireless SI path from the TX to the RX antenna is assumed to be linear. The purpose of the parallel SIC path is to achieve at least the SI cancellation for avoiding saturation of the low-noise amplifier (LNA) and the A/D converter in the receiver. The task of a neural network in the system thus would be the representation of an a-priori unknown nonlinear model function $f(s)$ of the PA and a linear impulse response model $w_k$ of the wireless SI path by means of suitable weights. The network output is then subtracted via D/A conversion and RF circuitry \cite{Duarte_2012,Valkama_2018} from the analog RX input $y_\text{H}(t)$ before its further propagation to the final output $r[k]$. The precise effect of the aforementioned RF circuitry will be neglected in our data. Naturally, any passive or active shielding between TX and RX sides of the system would relieve the digital SIC, but is beyond the scope of this paper also. The described system is meant as a context of our neural network study.

An opposite design with ''Wiener'' logic (i.e., generally a linear dynamical system element followed by memoryless nonlinearity) for SIC is shown in Fig.~\ref{fig:baseline}.b. The transmit signal $z(t)$ after the power amplifier is captured by attenuation (ATT) circuitry to bring the signal into the range of an auxiliary A/D converter. The conception here is that any PA nonlinearity is already encompassed by the utilization of the analog reference signal $z(t)$ as input for SIC and does not need to be modeled on the digital side \cite{Korpi_2014, Eltawil_2015}. A neural network would mimic the presumably linear TX to RX path of the system by means of an impulse response $w_k$ and the nonlinearity of LNA\&A/D with a subsequent nonlinear function representation $g(x)$ to attain SIC on the received signal $y_\text{W}[k]$ and the residual $r[k]$. \vspace{-1ex}

\section{Data Generation and Public Availability\!\!}
\vspace{-1ex}
\label{sec:data}

We consider the systems of Fig.~\ref{fig:baseline} for data relay and simulate high-throughput (HT) WLAN transmission \cite{Erceg_2004} in the complex baseband. Orthogonal frequency-division multiplex (OFDM) signals are generated for the ubiquitous IEEE-802.11 standard~\cite{Qureshi_2023}, specifically, the n-channel with $20\,\text{MHz}$ bandwidth, quadrature amplitude modulation with $64$ symbols, and coding rate of $5/6$.

Our SI channel assumes separate antennas for up- and downlink with TX and RX placed at a typical distance of $30$-$50$\,cm~\cite{Duarte_2013, Sethi_2014, Chen_2018}. The SI channel $h_\text{SI}[k]$ in the discrete-time domain thus can be represented as a multipath model with two components: a dominant, quasi-static internal SI channel $h_\text{iSI}[k]$ of the specific TX/RX antenna structure and an external SI channel $h_\text{eSI}[k]$ due to reflections from the environment, i.e., $h_\text{SI}[k] = h_\text{iSI}[k] + h_\text{eSI}[k]$.

\begin{figure}[t]
    \centering
    (a)\\
    \begin{minipage}[l]{\columnwidth}
        \def\svgwidth{\columnwidth} 
        \hspace{1ex}
        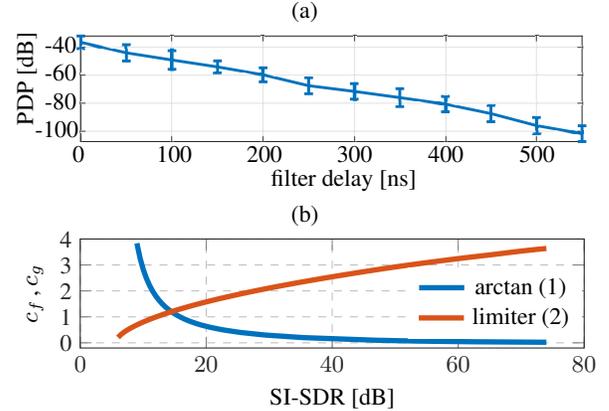
    \end{minipage}
    \\\vspace{1.5ex}
    (b)\\\vspace{-.5ex}
    \begin{minipage}[l]{\columnwidth}
        \hspace{3ex}
%
%
\begin{tikzpicture}

\definecolor{mycolor1}{rgb}{0.15,0.15,0.15}
\definecolor{mycolor2}{rgb}{0,0.447,0.741}
\definecolor{mycolor3}{rgb}{0.85,0.325,0.098}

\begin{axis}[%
scale only axis,
every outer x axis line/.append style={mycolor1},
every x tick label/.append style={font=\color{mycolor1}},
every outer y axis line/.append style={mycolor1},
every y tick label/.append style={font=\color{mycolor1}},
width=67mm,
height=14.5mm,
xmin=0, xmax=80,
ymin=-0.2, ymax=4,
ytick={0,1,2,3,4},
yticklabels={0,1,2,3,4},
xlabel={SI-SDR [dB]},
ylabel={$c_f$\,,\,$c_g$},
xmajorgrids,
ymajorgrids,
grid style={dashed},
legend style={at={(0.65,0.415)},anchor=west,draw=none}
]
\addplot[
color=mycolor2,
solid,
line width=2.0pt
]
coordinates{(9,3.83478) (9.5,3.26448) (10,2.83009) (10.5,2.48864) (11,2.21421) (11.5,1.98884) (12,1.80036) (12.5,1.6425) (13,1.50523) (13.5,1.38776) (14,1.28501) (14.5,1.19396) (15,1.11409) (15.5,1.04353) (16,0.978799) (16.5,0.920736) (17,0.868897) (17.5,0.820517) (18,0.777234) (18.5,0.737705) (19,0.698546) (19.5,0.667021) (20,0.635497) (20.5,0.603972) (21,0.57869) (21.5,0.554308) (22,0.529926) (22.5,0.505544) (23,0.486005) (23.5,0.467891) (24,0.449777) (24.5,0.431663) (25,0.413549) (25.5,0.396794) (26,0.384072) (26.5,0.371349) (27,0.358627) (27.5,0.345905) (28,0.333182) (28.5,0.32046) (29,0.307738) (29.5,0.296794) (30,0.288611) (30.5,0.280428) (31,0.272245) (31.5,0.264062) (32,0.255879) (32.5,0.247696) (33,0.239513) (33.5,0.23133) (34,0.223147) (34.5,0.214964) (35,0.206781) (35.5,0.199247) (36,0.194851) (36.5,0.190455) (37,0.186059) (37.5,0.181664) (38,0.177268) (38.5,0.172872) (39,0.168476) (39.5,0.16408) (40,0.159685) (40.5,0.155289) (41,0.150893) (41.5,0.146497) (42,0.142101) (42.5,0.137706) (43,0.13331) (43.5,0.128914) (44,0.124518) (44.5,0.120123) (45,0.115727) (45.5,0.111331) (46,0.106935) (46.5,0.102539) (47,0.0988855) (47.5,0.0962463) (48,0.0936072) (48.5,0.090968) (49,0.0883289) (49.5,0.0856897) (50,0.0830505) (50.5,0.0804114) (51,0.0782872) (51.5,0.0762581) (52,0.0742291) (52.5,0.0722) (53,0.0701709) (53.5,0.0681419) (54,0.0661128) (54.5,0.0640837) (55,0.0620547) (55.5,0.0600256) (56,0.0585881) (56.5,0.0571582) (57,0.0557283) (57.5,0.0542985) (58,0.0528686) (58.5,0.0514387) (59,0.0500088) (59.5,0.0485789) (60,0.047149) (60.5,0.0457191) (61,0.0442892) (61.5,0.0428593) (62,0.0414294) (62.5,0.0399997) (63,0.0391671) (63.5,0.0383346) (64,0.037502) (64.5,0.0366694) (65,0.0358368) (65.5,0.0350042) (66,0.0341716) (66.5,0.0333391) (67,0.0325065) (67.5,0.0316739) (68,0.0308413) (68.5,0.0300087) (69,0.0291761) (69.5,0.0283436) (70,0.027511) (70.5,0.0266784) (71,0.0258458) (71.5,0.0250132) (72,0.0241807) (72.5,0.0233481) (73,0.0225155) (73.5,0.0216829) (74,0.0208503)};
\addlegendentry{arctan \eqref{eq:NonlinPA}}

\addplot [
color=mycolor3,
solid,
line width=2.0pt
]
coordinates{(6,0.206573) (6.5,0.325365) (7,0.417091) (7.5,0.495419) (8,0.565664) (8.5,0.630161) (9,0.690328) (9.5,0.746981) (10,0.800791) (10.5,0.852012) (11,0.901108) (11.5,0.948253) (12,0.993623) (12.5,1.0375) (13,1.08004) (13.5,1.12137) (14,1.16159) (14.5,1.2008) (15,1.23874) (15.5,1.2758) (16,1.3121) (16.5,1.34764) (17,1.38216) (17.5,1.41611) (18,1.44952) (18.5,1.48196) (19,1.51398) (19.5,1.54549) (20,1.57624) (20.5,1.60668) (21,1.63647) (21.5,1.66582) (22,1.69481) (22.5,1.72319) (23,1.75139) (23.5,1.77892) (24,1.8063) (24.5,1.83316) (25,1.85979) (25.5,1.88603) (26,1.912) (26.5,1.93765) (27,1.96301) (27.5,1.98809) (28,2.01289) (28.5,2.03743) (29,2.06172) (29.5,2.08573) (30,2.10954) (30.5,2.13305) (31,2.15642) (31.5,2.17944) (32,2.20241) (32.5,2.22499) (33,2.24756) (33.5,2.26978) (34,2.29196) (34.5,2.31387) (35,2.33564) (35.5,2.35727) (36,2.37864) (36.5,2.4) (37,2.42095) (37.5,2.44191) (38,2.46266) (38.5,2.48327) (39,2.50382) (39.5,2.52409) (40,2.54436) (40.5,2.56442) (41,2.5844) (41.5,2.60431) (42,2.62399) (42.5,2.64367) (43,2.66312) (43.5,2.68245) (44,2.70177) (44.5,2.72087) (45,2.73997) (45.5,2.75895) (46,2.7778) (46.5,2.79666) (47,2.81515) (47.5,2.83357) (48,2.85194) (48.5,2.86997) (49,2.88799) (49.5,2.90585) (50,2.92336) (50.5,2.94088) (51,2.95812) (51.5,2.97505) (52,2.99198) (52.5,3.00865) (53,3.02508) (53.5,3.04151) (54,3.05769) (54.5,3.07363) (55,3.08956) (55.5,3.10541) (56,3.12109) (56.5,3.13676) (57,3.15243) (57.5,3.168) (58,3.18357) (58.5,3.19915) (59,3.21431) (59.5,3.22944) (60,3.24458) (60.5,3.25924) (61,3.27364) (61.5,3.28804) (62,3.3024) (62.5,3.3166) (63,3.3308) (63.5,3.345) (64,3.35914) (64.5,3.37325) (65,3.38735) (65.5,3.40153) (66,3.41633) (66.5,3.43112) (67,3.44592) (67.5,3.46098) (68,3.47615) (68.5,3.49132) (69,3.50605) (69.5,3.5202) (70,3.53435) (70.5,3.5485) (71,3.56163) (71.5,3.57463) (72,3.58763) (72.5,3.60057) (73,3.6123) (73.5,3.62403) (74,3.63575)};
\addlegendentry{limiter \eqref{eq:NonlinLNA}}

\end{axis}
\end{tikzpicture}
    \end{minipage}
    \vspace{-2ex}
    \caption{Facts of the data generation: (a) PDP of SI channels $h_\text{SI}[k]$ with standard deviations; (b) Parameters $c_f$ and $c_g$ vs.\ the SI-SDR.}
    \vspace{-2ex}
\label{fig:PDP_ParC_SISDR}
\end{figure}

We simulate $h_\text{SI}[k]$ based on measurements in~\cite{Chen_2018}, where real SI channels of several environments were characterized. Specifically, we rely on two observations regardless of the environment: (a) the power of the internal path is $[5; 10]$\,dB higher than the strongest external path and
(b) the root mean-square (RMS) delay spread is in the range of $[20; 40]\,$ns. We then rely on the WLAN multipath fading 'Model C' proposed by the HT Task Group for n-channel (TGn) waveform propagation \cite{Erceg_2004} and modify it\footnote{Note that both the HT baseband transmission of 802.11n and the generation of $h_\text{SI}[k]$ are simulated using the WLAN Toolbox of MATLAB~\cite{Cho_2010}.} for generation of our $h_\text{SI}[k]$. The resulting power delay profile (PDP) is illustrated in~\figref{fig:PDP_ParC_SISDR}.a and exhibits a duration of 550\,ns (corresponding to 12 samples at 20\,MHz sampling) to reach the noise floor.

\textit{Nonlinearities} of PA and LNA\&A/D in \figref{fig:baseline} are assumed to be memoryless~\cite{Eltawil_2015} with only amplitude-to-amplitude (AM/AM) distortion~\cite{Joung_2014}. The respective nonlinear functions are thus given as
\begin{align}
    \mathrm{PA}(s) &= F(|s|) \cdot e^{j \cdot \arg{(s)}}
    \,\,\text{with}\,\,
    F(|s|) = \text{arctan}(c_f \cdot |s|) \, ,
    \label{eq:NonlinPA}\\
    \!\!\!\mathrm{AD}(x) &= G(|x|) \cdot e^{j \cdot \arg{(x)}}
    \,\,\text{with}\,\,
    G(|x|) =
    \begin{cases}
        |x|,   &\!\!\!\! |x| < c_g \, , \\
         c_g,  &\!\!\!\! |x| \geq c_g \, ,
    \end{cases}
    \label{eq:NonlinLNA}
\end{align}
where $c_f$ and $c_g$ are scale and saturation parameters, respectively. The PA function intuitively refers to a soft limitation of the power amplifier while the LNA function refers to the range of the related A/D converter. The related distortions can be quantified by the scale-invariant signal-to-distortion ratio (SI-SDR)~\cite{LeRoux_2019sisdr}.  \figref{fig:PDP_ParC_SISDR}.b shows the dependency with $c_f$ and $c_g$ parameters.

The SIs of system options in \figref{fig:baseline} are then generated at discrete time $k$ for the ''Hammerstein'' (H) system with PA nonlinearity as
    \begin{equation}
        y_{\text{H}}[k] = \mathrm{PA}(\,s[k]\,) * h_\text{SI}[k] + n[k] \,
        \label{eq:Hammerstein}
    \end{equation}
and for the ''Wiener'' (W) system with LNA\&A/D nonlinearity as
    \begin{equation}
        y_{\text{W}}[k] = \mathrm{AD}(\,z[k] * h_\text{SI}[k] + n[k] \,) \, ,
        \label{eq:Wiener}
    \end{equation}
where $*$ denotes discrete-time convolution and $n[k]$ the remote signals and receiver noise at -90\,dB below the SI.

\textit{Three datasets} are then generated for ''Hammerstein'' (H) and ''Wiener'' (W) systems each, i.e., with variant or invariant SI channel $h_\text{SI}[k]$ and with variant or invariant nonlinearity (NL), labeled as
\begin{enumerate}
    \item[(H)] ''invNL+invSI'', ''invNL+varSI'', and ''varNL+varSI''
    \item[(W)] ''invSI+invNL'', ''varSI+invNL'', and ''varSI+varNL''\,.
\end{enumerate}
Each of the dataset consists of 10 file IDs each representing a time-invariant SI system and a WLAN packet size of 3218 discrete-time samples\footnote{The generated datasets and the MATLAB code for generation are available on GitHub under \href{https://github.com/STHLabUOL/SICforIBFD}{https://github.com/STHLabUOL/SICforIBFD}.}. The identifier ''inv'' here refers to invariable SI channels or nonlinearities across the data set, while the label ''var'' means a variable SI channel for each file ID based on new path gains for $h_\text{SI}[k]$ or a variable nonlinearity with corresponding $\text{SI-SDR}$ uniformly drawn from the interval $\text{SI-SDR}_0\pm 4\,\mathrm{dB}$. Note that the data sets are drawn from two different Matlab scripts specifically developed for the (H) and (W) systems options from~\figref{fig:baseline}.

\section{Neural-Network Architectures}
\vspace{-1ex}
\label{sec:architectures}

We present a range of models to cope with the data according to the Hammerstein system of Fig.~\ref{fig:baseline}.a. All models implement complex-valued computations for the complex baseband and are well guided by the given arrangement of the Hammerstein system. The model sizes are deliberately small in terms of the number of trainable weights. Our models may thus be considered as baseline architectures, but since the amount of data available for system identification problems is typically moderate, the small model sizes are perfectly appropriate. The idea of system modeling is enforced by the single-input/single-output architectural property of all networks in parallel to the Hammerstein system. The complex-valued residual $r[k]\!=\!y_\text{H}[k]\!-\!\widehat{y}_\text{H}[k]$ between system and model is to be minimized and, thereto, we employ the mean-square error (MSE) loss.

\begin{figure}[!tb]
  \centering 
  \includegraphics[width=0.72\columnwidth]{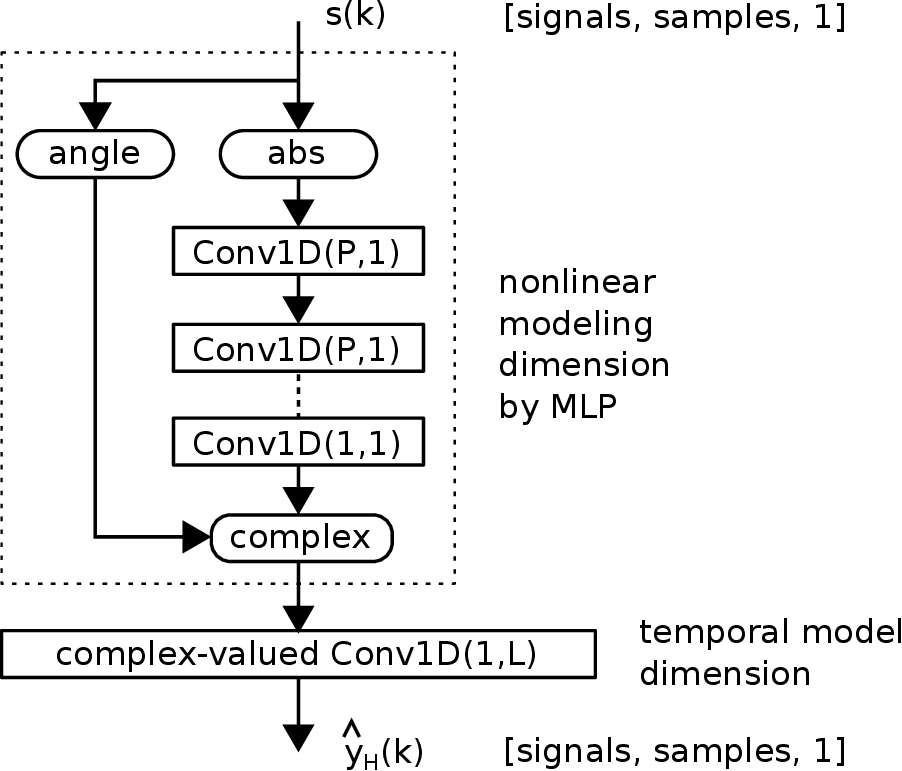}
  \caption{Global complex-valued Hammerstein model.}
  \vspace{0.75ex}
  \label{fig:modelPlain}
\end{figure}

Fig.~\ref{fig:modelPlain} shows a first model with complex-valued input signal $s[k]$, where the input tensor hosts the file IDs available for optimization as the ''signals'' in its batch dimension and the corresponding temporal ''samples'' of each signal in its first feature dimension. The last dimension merely refers to a single input channel. Cartesian $s[k]$ is then split into magnitude and phase to resemble a physically motivated nonlinear PA in the magnitude~\cite{Joung_2014, Tellado_2003}. Nonlinear modeling $f(s)$ on the magnitude is here represented by a multilayer perceptron (MLP) where the number of units per layer is here implemented by the number of filters $P$ of convolutional layers. Each time step of the input is treated independently, i.e., the MLP contributes a memoryless nonlinear modeling to the system. Its single output unit is then recombined with the original phase. The tensor format at this point of the model complies with the input tensor. In the final model stage, the temporal samples are taken to a complex-valued linear convolution with filter length $L$ to represent the dynamical system part $w_k$ of the underlying Hammerstein plant in Fig.~\ref{fig:baseline}.a. 
 
\begin{figure}[!htb]
  \centering 
  \includegraphics[width=0.72\columnwidth]{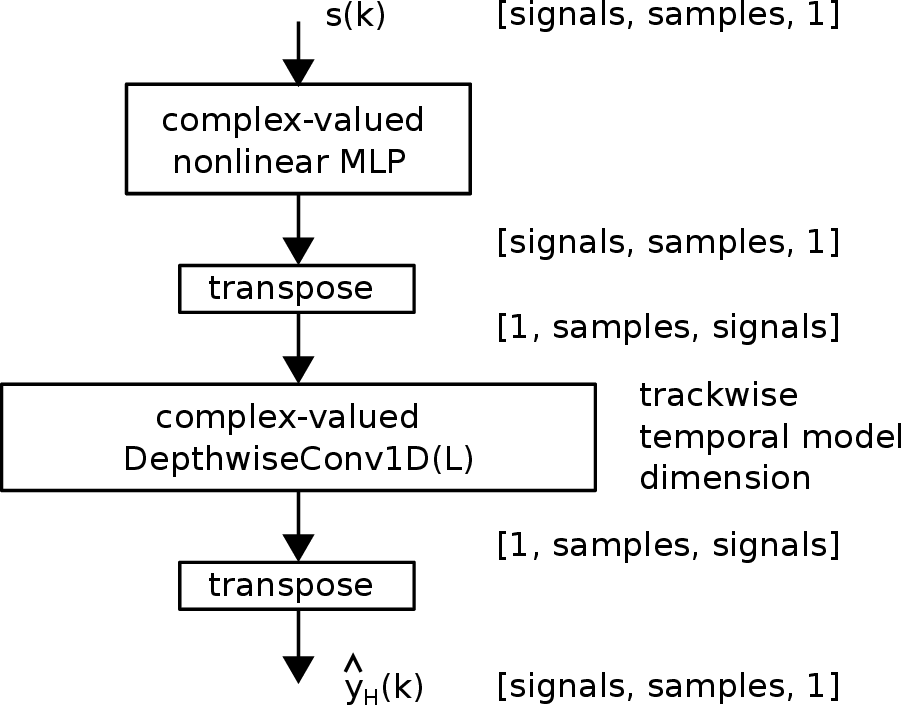}
  \caption{Adaptive complex-valued Hammerstein model.}
  \vspace{0.75ex}
  \label{fig:modelAdaptive}
\end{figure}

The first model architecture seeks an average (global) Hammerstein representation of the signals available for optimization. This approach is likely to fail the successful fitting of training data in the presence of variable SI channels in the underlying Hammerstein data. Our next model in Fig.~\ref{fig:modelAdaptive} therefore reuses the complex-valued MLP stage for nonlinear representation, but a follow-up transposition then takes the different signals with possibly different SI channels to the last dimension of the tensor. From here, a ''depthwise'' convolution (which is Tensorflow jargon and assumes the described configuration of its input tensor) applies its competence of individual kernels to the individual signals in order to represent their possibly individual SI channel. The size of the last dimension is preserved by this operation and the output tensor is finally reverse transposed to comply with the target signals for loss computation.

\begin{figure}[!htb]
  \centering 
  \includegraphics[width=0.72\columnwidth]{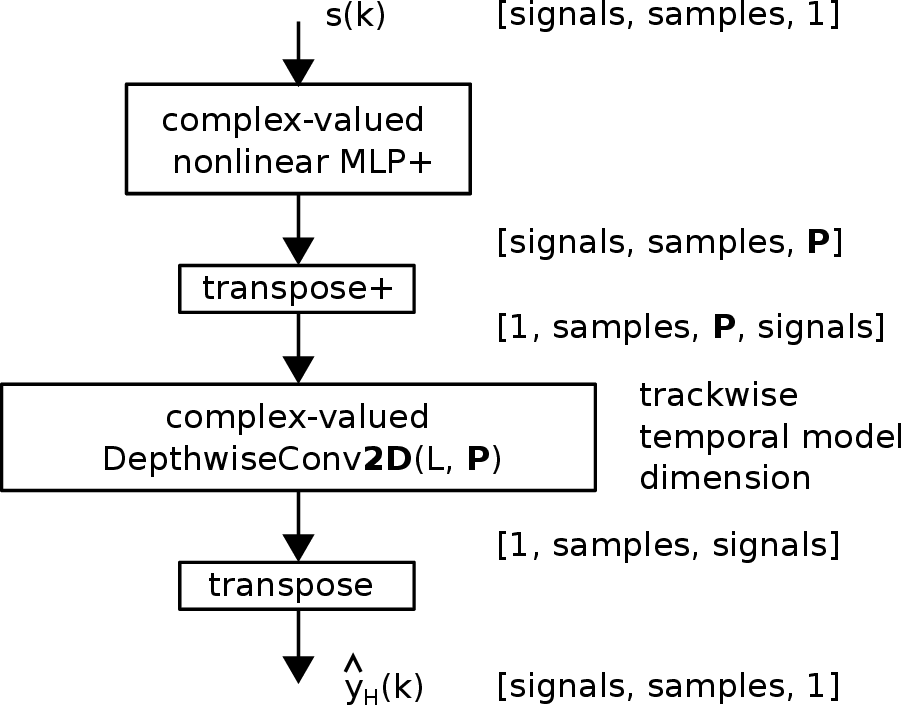}
  \caption{Parallel adaptive complex-valued Hammerstein model.}
  \vspace{-2ex}
  \label{fig:modelParallel}
\end{figure}

The last model in Fig.~\ref{fig:modelParallel} applies a generalization of the nonlinear MLP stage while retaining the depthwise treatment of the second model (to maintain its ''adaptivity'' to different signals). Our nonlinear generalization adopts an architectural property of, e.g., memory polynomials \cite{Morgan2006}, for network design. Specifically, the MLP now prepares multiple nonlinear output units, i.e., expanding its single input to $P$ output channels. With the transposition of signals to the last dimension, as before, the MLP output allows $P$ different depthwise convolutional kernels to form the final model output. This operation is implemented by 2D convolution and it swallows the $P$ nonlinear units before reverse transposition to familiar output format.

\vspace{-1ex}
\section{Training and Evaluation}
\vspace{-1ex}
\label{sec:results}

With the reproducible data of Sec.~\ref{sec:data}, we here provide an evaluation of the range of neural network models of Sec.~\ref{sec:architectures} on the Hammerstein subset of the data. For further orientation we also include results of the complex-valued memory polynomial \cite{Morgan2006}. The available data and baselines would hopefully be useful for rapid setup and coherent comparison of further models for SI system modeling.

The MLP of the neural network models consist of~2 hidden and~1 output layer with nonlinear order (i.e., the number of hidden units) P=8 and thus 8+64+8=80 trainable weights. The parallel Hammerstein model discards the output layer in order to provide multiple output units for expansive nonlinear modeling. The linear model order (i.e., the kernelsize) is chosen as L=32 which refers to a bit over-modeling of the data. We thus have 32 additional trainable weights for global Hammerstein, 10$\cdot$32=320 additional trainable weights for adaptive Hammerstein, and 8$\cdot$320=2560 additional trainable weights for parallel Hammerstein modeling. Despite the performance limitations to be seen in the following, we can report that a pure upscaling of the model sizes has not delivered stronger results. 

Our implementation and training of the models uses Tensorflow with the Keras backend. In each of the following experiments the training uses different subsets of the data with increasing levels of difficulty. All data samples of a specific experiment together form a single batch for model optimization. The minimization of the MSE loss is pursued until convergence with a fixed number of $10^4$ epochs in order to study the related learning behavior. We rely on the Adam optimizer with learning rate of 0.01 in all cases.   

\begin{figure}[!tb]
  \centering 
  \includegraphics[width=0.99\columnwidth]{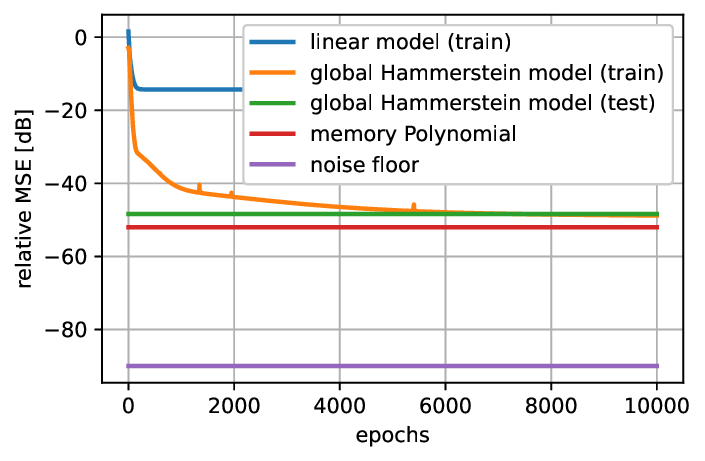}
  \vspace{-3ex}
  \caption{System identification on invariant Hammerstein data.}
  \vspace{-1ex}
  \label{fig:invariantHammersteinData}
\end{figure}

Fig.~\ref{fig:invariantHammersteinData} shows the MSE of SI modeling for data created by an invariant Hammerstein SI system. With a linear model on this nonlinear data, the MSE is limited to -13\,dB which is plausible with our SI-SDR of 10\,dB for this analysis. The global Hammerstein model (Fig.~\ref{fig:modelPlain}) clearly better fits the nonlinear data and attains training MSE of about -50\,dB. The model is validated to about the same MSE with test data constructed with different waveforms on the same Hammerstein system. However, the neural model cannot reach to the noise floor of -90\,dB, nor does the memory polynomial.

\begin{figure}[!tb]
  \centering 
  \includegraphics[width=0.99\columnwidth]{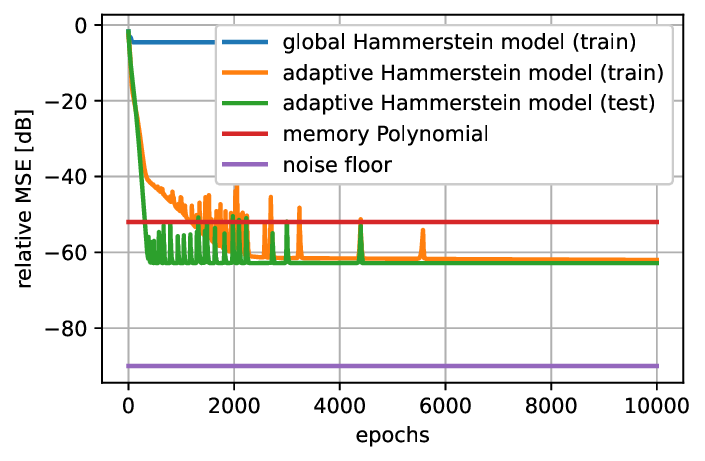}
  \vspace{-3ex}
  \caption{Data with invariant nonlinearity and variable SI channel.}
  \vspace{-1ex}
  \label{fig:invariantNLvariantRIR}
\end{figure}

Fig.~\ref{fig:invariantNLvariantRIR} steps up to Hammerstein SI created with variable SI channels. Here a global Hammerstein model already fails to fit the training data, which is plausible from the perspective of nonlinear system identification on inconsistent plants (the global Hammerstein model supposedly represents an average of different SI channels). Our adaptive Hammerstein model (Fig.~\ref{fig:modelAdaptive}) was constructed for the case at hand to represent with its individual kernels the different SI channels and, in this way, to fit the inconsistent data. An MSE of about -60\,dB is here restored successfully by the adaptive Hammerstein model with training and testing data, where the test data consists in different waveforms, SI channels, and noise.  

\begin{figure}[!tb]
  \centering 
  \includegraphics[width=0.99\columnwidth]{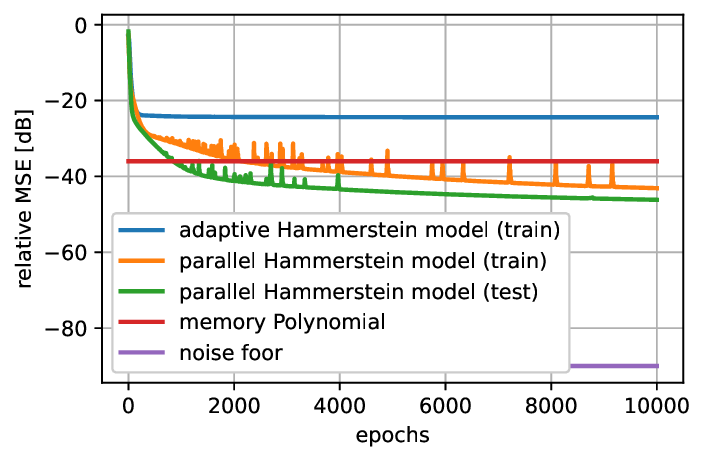}
  \vspace{-3ex}
  \caption{Data with variable nonlinearity and variable SI channel.}
  \vspace{-1ex}
  \label{fig:variantNLvariantRIR}
\end{figure}

Fig.~\ref{fig:variantNLvariantRIR} then uses Hammerstein data with variable SI channels and variable nonlinearities per sample which is supposed to further trouble system modeling by neural networks. Our before successful adaptive Hammerstein model here indeed demonstrates a limitation of MSE to -25\,dB, but the parallel Hammerstein model (Fig.~\ref{fig:modelParallel}) can almost restore the -50\,dB in its capacity of expanding different nonlinear functions by means of individual kernels on multiple MLP output units. The set of weights shared between training and testing is in this case narrowed down to just the hidden layers of the MLP. Notably, the memory polynomial model in this case falls behind to about -35\,dB which we were able to attribute to the dominant effect of the lower SI-SDR values around mean $\text{SI-SDR}_0$ of 10\,dB.

\begin{figure}[!tb]
  \centering 
  \includegraphics[width=0.99\columnwidth]{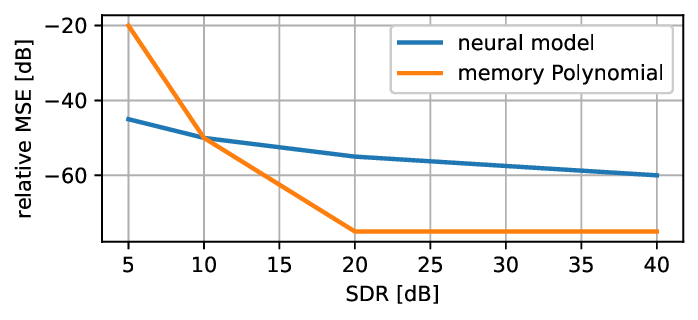}
  \vspace{-3ex}
  \caption{Neural modeling and memory polynomial vs.\ SDR.}
  \vspace{-7ex}
  \label{fig:SDRfunction}
\end{figure}

This last observation motivates the analysis of Fig.~\ref{fig:SDRfunction}, where the mean MSEs of our neural modeling and of the memory polynomial baseline are compared for different SDR. Here in turns out a mediocre, while robust functional modeling ability of the neural representation, i.e., with relatively weak dependence on the SDR of the data. The memory polynomial degrades with low SDR.

\vspace{-1ex}
\section{Conclusion}
\vspace{-1ex}
\label{sec:conclusion}

Our neural representation of RF self-interference (SI) shows that models architected with a strong orientation in domain knowledge of SI systems may encounter various difficulties depending on the data. Specifically, we must provide models with a sense of adaptivity to address the variability of the SI channel or the nonlinear function. The SDR of the nonlinearity is, however, not a critical factor, as opposed to the established memory polynomial. The neural MSE of about -50\,dB for the data at hand would be a good contribution, but will not be sufficient for SI cancellation to reach the noise level. We encourage the study of further models for SI representation.

\bibliographystyle{IEEEbib}
\bibliography{refs}

\end{document}